\documentclass[showpacs,twocolumn,floatfix,aps,
]{revtex4}
\usepackage{epsfig}
\graphicspath{{figures/}}

\newcommand{\beq}{\begin{equation}}  
\newcommand{\eeq}{\end{equation}}  
\newcommand{\bea}{\begin{eqnarray}}  
\newcommand{\eea}{\end{eqnarray}}  
\newcommand{\e}[1]{\cdot 10^{#1}} 	
\newcommand{\bv}[1]{\mathbf{#1}} 	
\newcommand{\eqref}[1]{(\ref{#1})}

\begin{document}

\title{Intermittent boundary layers and torque maxima in Taylor-Couette flow}
\author{Hannes J. Brauckmann and Bruno Eckhardt}
\affiliation{Fachbereich Physik, Philipps-Universit\"at Marburg, 
D-35032 Marburg, Germany}
\affiliation{J.M. Burgerscentrum, TU Delft, Delft, The Netherlands}
\date{\today}

\begin{abstract} 
Turbulent Taylor-Couette flow between counter-rotating cylinders develops
intermittently fluctuating boundary layers for sufficient counter-rotation.
We demonstrate the phenomenon in direct numerical simulations for radius ratios $\eta=0.5$
and $0.71$ and propose a theoretical model for the critical value in 
the rotation ratio. Numerical results as well as experiments show that the
onset of this intermittency coincides with the maximum in torque. 
The variations in torque correlate with the variations in mean Taylor 
vortex flow which is first enhanced for weak counter-rotation,
and then reduced as intermittency sets in. 
To support the model, we compare to numerical results, 
experiments at higher Reynolds numbers, and to Wendt's data.
\end{abstract}


\maketitle

\section{Introduction}
The flow between concentric cylinders has served as a paradigm for 
the transition to turbulence since Taylors 1923 characterization of the
bifurcation from laminar to vortical flows \cite{Taylor1923}. Many 
transitions between spatially and temporally simple flow states like 
vortices, modulated vortices and traveling waves, are accessible by standard
bifurcation theory and have been described and studied in considerable detail
\cite{Koschmieder1993,Andereck1986}.
The turbulent states that are reached after several of these bifurcations
are not always homogeneous but can show patchiness in the form of
turbulent spots or turbulent spirals \cite{Coles1965,VanAtta1966,Meseguer2009b,Dong2011}. 
In addition to such azimuthal and axial modulations, 
Coughlin and Marcus \cite{Coughlin1996} have described a radial
inhomogeneity for counter-rotating cylinders, 
which consists of turbulent bursts that have been seen in experiments by Colovas and Andereck \cite{Colovas1997}.
Measurements in the Twente Taylor-Couette facility have confirmed the presence of this inhomogeneity
up to Reynolds numbers of $10^6$ \cite{vanGils2012}. 
Moreover, the onset of this instability has been linked to the maximum in torque 
that appears for moderate counter-rotation \cite{vanGils2012,Paoletti2011}.

Coughlin and Marcus \cite{Coughlin1996} already suggested that this radial 
inhomogeneity should be linked to the presence of a neutral surface of vanishing angular
velocity in the laminar Couette profile for counter-rotating cylinders. 
The region between the inner cylinder and the neutral surface is inviscidly
unstable by the Rayleigh criterion, whereas the region between the
neutral surface and the outer cylinder is Rayleigh stable. These 
considerations do not provide an immediate prediction for the onset of 
inhomogeneity since the neutral surface appears with infinitesimal amounts 
of counter-rotation already. 
We here discuss the extensions needed in order 
to derive a predictive theory for the onset of this intermittency, compare
it with observations for different radius ratios, and describe the link to 
the torque maxima.

The outline of the paper is as follows. In section II we introduce the numerical
simulations and describe the phenomenon. We also identify the onset of intermittency and 
summarize numerical and experimental results  for torque maxima, including a reanalysis of 
the data of Wendt \cite{Wendt1933} which is documented in appendix \ref{sec:wendt}. 
In section \ref{sec:theory} we present the argument for the boundary layer intermittency, 
and in section \ref{sec:LSC} we describe the link to the torque maximum.
We conclude with a few remarks in section \ref{sec:remarks}.

\section{Boundary layer intermittency}
\label{sec:BLfluct}
In order to introduce the phenomenon we present results from direct numerical simulations for different
radius ratios. We solve the incompressible Navier-Stokes equation 
with the spectral scheme explained in \cite{Meseguer2007}. For the dimensionless units
we measure all lengths and times in 
units of the gap width $d=r_o-r_i$, where $r_o$ and $r_i$ are the radii of the outer and inner 
cylinders, respectively, and the viscous time $d^2/\nu$. To avoid the end-effects caused by top and bottom lids in experiments, an additional periodicity in axial direction of length $L_z$ is introduced resulting in an aspect ratio $\Gamma=L_z/d$. Here $\Gamma=2$, which allows for one Taylor vortex pair when the outer cylinder is at rest. Fourier modes and Chebyshev polynomials are employed for expansions in the two periodic and  the wall-normal direction, respectively. We simulate a domain of reduced azimuthal length, i.e. one third for \mbox{$\eta=r_i/r_o=0.5$} and one ninth for $\eta=0.71$ of the full azimuthal length. Consequently the flow field repeats three (nine) times to fill the entire circumference. We tested that the shorter azimuthal period does not influence the computed torques. The criteria used to test and verify the code
are detailed in \cite{Brauckmann}. Specifically, the 
spatial resolution, characterized by the number of modes $(N_z,N_\varphi,N_r)$ in each direction, is chosen so that three convergence criteria are satisfied: torque computed at the inner and outer cylinder has to agree within a relative deviation of 
$5\e{3}$, the expansion coefficients in each direction have to cover a range of at least $10^4$ and the energy dissipation 
estimated from the torque has to agree with the volume energy dissipation rate to within $10^{-2}$. All these requirements are met in all simulations shown here.

We consider radius ratios \mbox{$\eta=0.5$} and \mbox{$\eta=0.71$}. 
In order to assess the influence of the mean system rotation on turbulent characteristics, 
the simulations are performed at a fixed shear between the cylinder walls,  defined by 
Dubrulle et al. \cite{Dubrulle2005} as
\begin{equation}
	Re_S=\frac{2}{1+\eta}\, \left|\eta\,Re_o-Re_i \right| \,,
\end{equation}
with $Re_i=(r_o-r_i)r_i\omega_i/\nu$ and $Re_o=(r_o-r_i)r_o\omega_o/\nu$. Here, $\omega_i$ and $\omega_o$ denote the angular velocity of the inner and outer cylinder and $\nu$ the kinematic viscosity. For both radius ratios we realize various mean rotations characterized by the rotation ratio $\mu=\omega_o/\omega_i$ for the same shear \mbox{$Re_S=2.0\e{4}$}. According to Lathrop et al. \cite{Lathrop1992}
this value is high enough that the vortices noticeable for lower \mbox{Re} have disappeared for $\mu=0$.
The dimensionless torque $G$ exerted on the inner and outer cylinder is obtained from  
\begin{equation}
  G(t)=\nu^{-2}J^\omega=\nu^{-2}r^3\left(\left<u_r\omega\right>_{A(r)}-\nu \partial_r\left<\omega\right>_{A(r)} \right) \, ,
	\label{eq:Jw}
\end{equation}
where $u_r$ and $\omega=u_\varphi/r$ denote the radial and angular velocity, respectively, and $\left<\cdots\right>_{A(r)}$ stands for an area average over the surface of a concentric cylinder \cite{Eckhardt2007}. The long-time mean value is obtained
from an additional average over time.

\begin{figure} 
  \centering
  \includegraphics{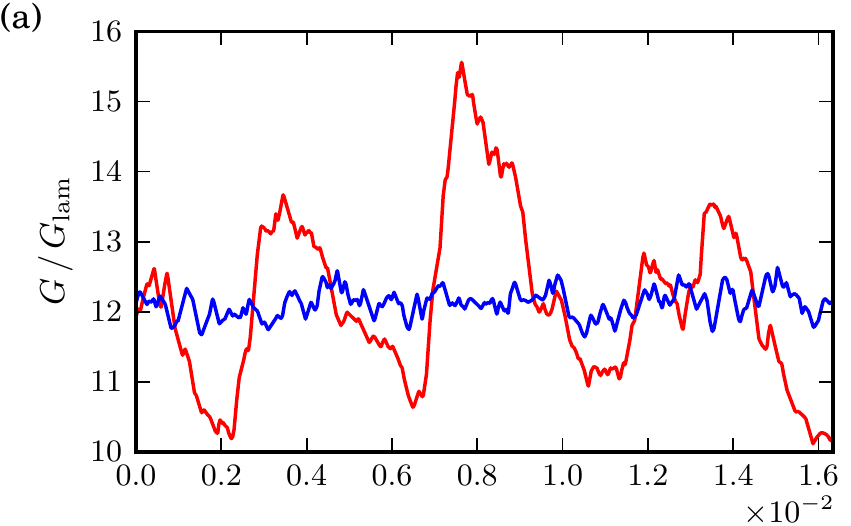}		
	\includegraphics{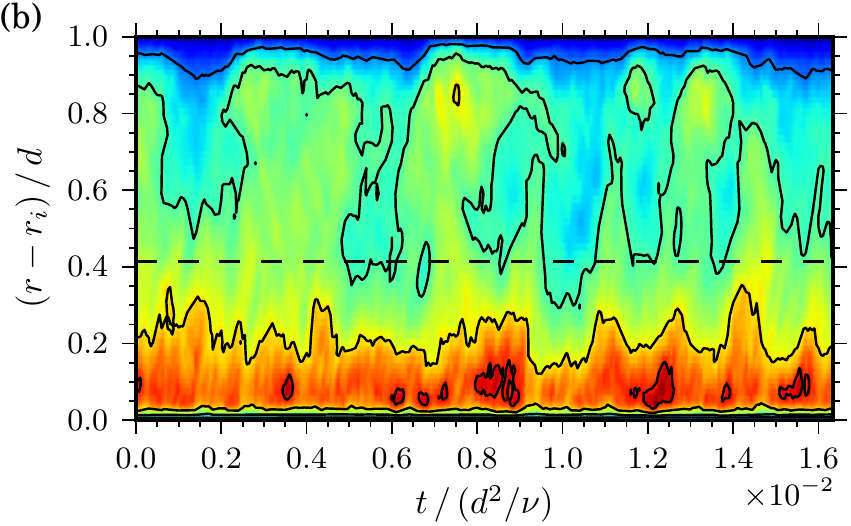}		
 \caption[]{Turbulent bursts for $\eta=0.5$, $Re_S=20000$ and $\mu=-0.5$. (a) Time-series of torque at the inner cylinder (blue) and the outer cylinder (red) divided by its laminar value. (b) Spatio-temporal plot of cross-flow energy, $\left<u_r^2+u_z^2\right>_{A(r)}$, averaged along the periodic coordinates with red (blue) indicating high (low) energy. The dashed line indicates the laminar neutral surface.
  \label{fig:tseries}}
\end{figure}

For co-rotation, i.e. $\mu\geq0$ , torque values at the inner and outer cylinders agree on average and in their
fluctuations \cite{Brauckmann}. However, the situation drastically changes for strong counter-rotation as illustrated by the torque time-series for $\mu=-0.5$ in Fig. \ref{fig:tseries}(a). Near the inner cylinder, fluctuations are small and without long time
variation. At the outer cylinder the torque exhibits fluctuations of relatively strong amplitude and slow dynamics that qualitatively differ from the ones at the inner cylinder. These slow fluctuations reflect an intermittent turbulent activity in the vicinity of the outer cylinder as demonstrated by the cross-flow energy in Fig. 1(b). The cross-flow energy, 
\begin{equation}
E_{cf}(r,t)=\left<u_r^2+u_z^2\right>_{A(r)}
\end{equation}
measures the energy content in the transverse velocity components at a radial distance $r$ and at an instant of time. 
Small values indicate a flow close to laminar, large values a turbulent state. One notes that near the inner cylinder
the flow is more or less homogeneously turbulent, whereas towards the outer cylinder one observes an increased activity
synchronized with the increase in mean torque. 

For this geometry, the Rayleigh-stable region of the laminar profile extends over the interval \mbox{$(r-r_i)/d\in[0.414,1]$}, with the lower bound marked by a dashed line in 
Fig. \ref{fig:tseries}(b).
The presence of the bursts shows that this stability is not maintained, but the position 
of the laminar line still marks the
transition between small radial variations near the inner cylinder and larger ones towards the
outer cylinder.

For \mbox{$Re_S\sim2.3\e{3}$} Coughlin and Marcus \cite{Coughlin1996} described the bursts as instability of spatially ordered 'interpenetrating spiral' flow. We here observe turbulent flow in the Rayleigh unstable inner region at $Re_S=2.0\e{4}$ that is accompanied by intermittent turbulence in the outer one. The connections to spirals cannot be followed up because of the
limited radial and azimuthal domain size that can be computed. The experiments of \cite{vanGils2012} extend this 
observation to even larger Reynolds numbers $Re_S\sim10^6$ where the presence of turbulent bursts in the outer region
is deduced from a bimodal distributions of the angular velocity.

\begin{figure} 
  \centering
  \includegraphics{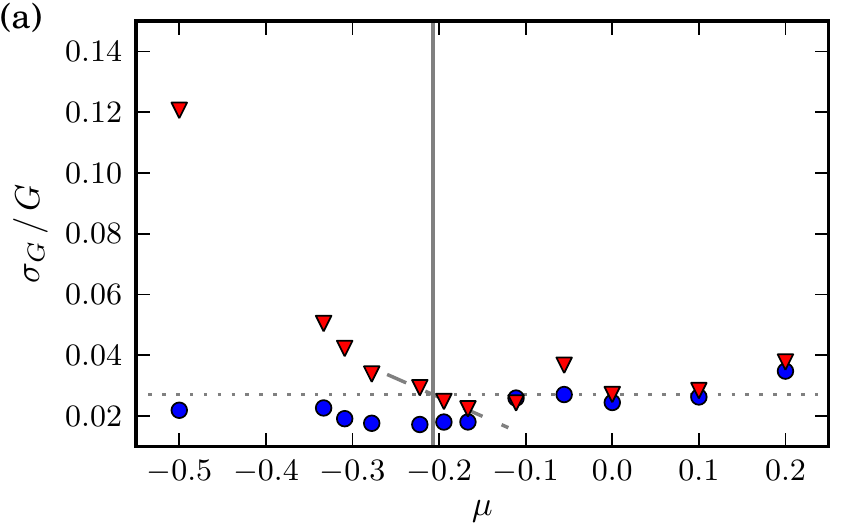}		
  \includegraphics{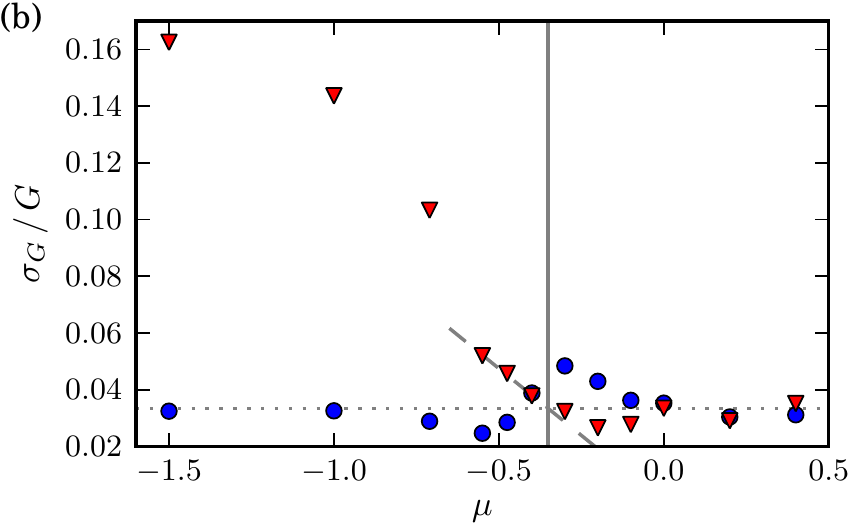}		
 \caption[]{Standard deviation $\sigma_G$ of temporal torque fluctuations (divided by the mean $G$) computed at the inner cylinder (blue circles) and the outer cylinder (red triangles) for a constant shear and various global rotations. The dotted line marks the fluctuation base level, the dashed line a linear fit to the data and the solid line the intersection point $\mu_c$. (a) $\eta=0.5$ and $Re_S=20000$. (b) $\eta=0.71$ and $Re_S=19737$. 
  \label{fig:fluct}}
\end{figure}

The onset of the intermittent behavior is accompanied by an increase in the torque fluctuations. Therefore, 
we study the standard deviation $\sigma_G$ of the torque relative to the mean for the outer cylinder, i.e. the 
ratio $\sigma_G/G$, as an indicator and deduce a critical value $\mu_c(\eta)$ for the onset 
from the requirement that $\sigma_G/G$ exceeds the base 
level for $\mu=0$, cf. Fig. \ref{fig:fluct}. The choice of $\sigma_G/G$ as measure for the transition is supported by the observation that it remains relatively unaffected by variations of $\mu$ at the inner cylinder.  For the numerical simulations we find the critical values
\begin{eqnarray}
 \mu_c(0.5) &=& -0.207 \pm 0.014  \nonumber \\
 \mu_c(0.71) &=& -0.351 \pm 0.050 \,,
 \label{eq:mucrit}
\end{eqnarray}
marked by vertical lines in Fig. \ref{fig:fluct}. The uncertainties are estimated as half the gap in $\mu$ between the two data points
next to the maximum. The critical point varies with $\eta$, and we will return to this feature in section \ref{sec:theory}. The onset of intermittent bursts was determined experimentally at $\mu_c(0.716)\approx-0.368$ for $Re_S\sim10^6$ \cite{vanGils2012}. The differences between this value and our observation \eqref{eq:mucrit} may be due to our uncertainty in $\mu_c$ or the difference in Reynolds number, but they are not significant.

\begin{figure} 
  \centering
  \includegraphics{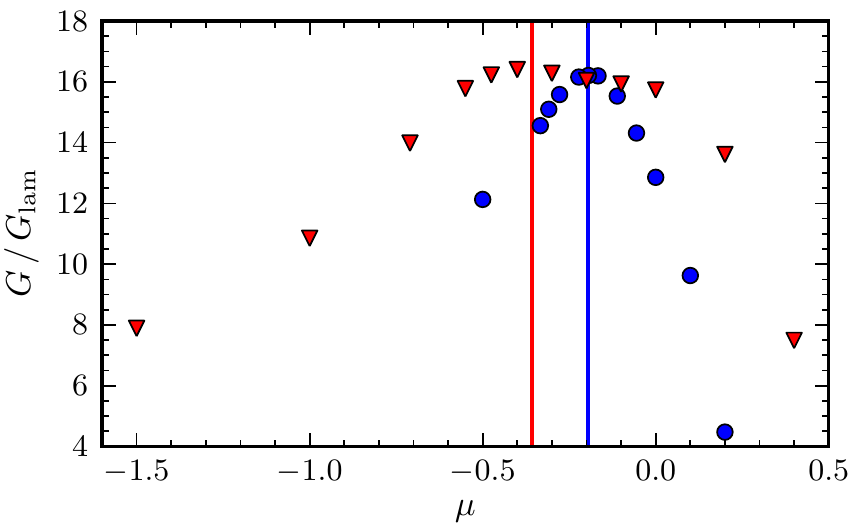}		
 \caption[]{Dependence of the total torque on the global rotation for $\eta=0.5$ and $Re_S=20000$ (blue circles) and for $\eta=0.71$ and $Re_S=19737$ (red triangles). Error bars based on the statistical uncertainty would be smaller than the symbols.
  \label{fig:nusselt}}
\end{figure}

The rotation ratio $\mu_{\mathrm{max}}(\eta)$ of maximal torque was identified independently in two experiments, \cite{vanGils2011,vanGils2012,Paoletti2011}. For a constant shear, they report a torque maximization for $\mu_{\mathrm{max}}(0.716)=-0.33 \pm 0.04$ \cite{vanGils2012} and for $\mu_{\mathrm{max}}(0.7245)=-0.333$ \cite{Paoletti2011}. Numerical simulations show that the torque maximum for counter-rotation only occurs after a shift of the maximizing $\mu$-value with increasing $Re_S$ \cite{Brauckmann}. Figure \ref{fig:nusselt} shows the computed torques for $Re_S=2.0\e{4}$ 
just at the beginning of the asymptotic regime.
We determine the rotation ratio of optimal transport as the maximum of a quadratic fit, $G/G_{\mathrm{lam}}=c_2\mu^2+c_1\mu+c_0$, to five data points and find 
\begin{eqnarray}
 \mu_{\mathrm{max}}(0.5) &=& -0.195 \pm 0.019 \nonumber \\
 \mu_{\mathrm{max}}(0.71) &=& -0.357 \pm 0.060 \, .
 \label{eq:mumax}
\end{eqnarray}
The uncertainties are deduced from the relative confidence interval $\Delta G/G$ which results from temporal torque fluctuations. This uncertainty in the torque values transforms into an uncertainty in the maximum location of the quadratic fit, i.e. 
\begin{equation}
  \Delta \mu_{\mathrm{max}}= \sqrt{-\frac{\Delta G}{c_2\,G} \, \left(G/G_{\mathrm{lam}}\right)_{\mathrm{max}}} \,
  \label{eq:uncert}
\end{equation}
with the fit coefficient $c_2$ and the maximal rescaled torque $(G/G_{\mathrm{lam}})_{\mathrm{max}}$.

For both radius ratios, the maximizing global rotation $\mu_{\mathrm{max}}$ compares well with the transition to the radial inhomogeneity at $\mu_c$, as it was previously observed experimentally by van Gils et al. \cite{vanGils2012} for $\eta=0.716$. On the other hand, the critical values for $\eta=0.71$ are more uncertain, which complicates the identification of a correspondence between them. 
We note that the torque maximization at $\mu_{\mathrm{max}}(0.71)=-0.357$ falls in line with the experimental 
observations $\mu_{\mathrm{max}}(0.716)=-0.33\pm0.04$ and $\mu_{\mathrm{max}}(0.7245)=-0.333$ \cite{vanGils2012,Paoletti2011}. 

A final data point is provided by Wendt's data \cite{Wendt1933}, 
which are reanalyzed in the way used in the recent experiments 
in appendix \ref{sec:wendt}.  Wendt worked with the radius ratio of $\eta=0.680$, and the maximum in his torque data lies near 
\begin{equation}
 \mu_{\mathrm{max}}(0.680)= -0.295 \pm 0.113 \, .
\end{equation}

\section{Onset of the radial inhomogeneity}
\label{sec:theory}
In order to connect the onset of fluctuations with the rotation ratio for maximal torque
and its dependence on the radius ratio, we ask the following questions: (i) What is the physical mechanism that determines the onset of fluctuations and can one derive a prediction for the onset from it? (ii) Why should the rotation ratio of torque maximization coincide with the onset of fluctuations?

A first answer to these questions is provided by van Gils et al. \cite{vanGils2012}: They suggest that the rotation ratio for maximal torque is determined by locations in parameter space $(Re_i,Re_o)$ that are equally distant from the Rayleigh stability lines $\mu=\eta^2$ and $\mu=-\infty$. This condition results in the so called ``angle bisector'',
\begin{equation}
  \mu_{\mathrm{bis}}(\eta)= \frac{-\eta}{\tan\left[\frac{\pi}{2} - \frac{1}{2}\arctan(\eta^{-1})\right]} \; ,
  \label{eq:bisector}
\end{equation}
for the location of the torque maximum \cite{vanGils2012}. 
Secondly, they argue that the onset of fluctuations has to coincide with \eqref{eq:bisector} since the intermittent behavior in the outer layer reduces radial transport of momentum the torque starts to drop. 
While the angle bisector agrees well with their measured optimum \mbox{$\mu_{\mathrm{max}}(0.716)=-0.33$}, it disagrees with our simulation result for $\eta=0.5$, since
\mbox{$\mu_{\mathrm{max}}(0.5)=-0.195$} whereas Eq. \eqref{eq:bisector} gives \mbox{$\mu_{\mathrm{bis}}(0.5)=-0.309$}.

We here propose an explanation for the onset of fluctuations that is not based on the stability of laminar flow but on properties of turbulent flows in general and turbulent Taylor vortices in particular. 
The key idea is that the turbulent flow detaches because the inner part that is driven by the Rayleigh-unstable region 
is not sufficiently strong to maintain persistent turbulence across 
the Rayleigh-stable region to the outer cylinder. But the outer region cannot return to
laminar for all times, because the turbulent transport and the friction has to be the same, independent of the radial distribution \cite{Eckhardt2007}. 
The radial range over which the inner unstable region can maintain a turbulence is somewhat larger than the inner 
Rayleigh-unstable region. The simple inviscid stability calculations for counter-rotation alluded to before 
gives a neutral surface at radius
\begin{equation}
 r_n(\mu) =r_i\,\sqrt{\frac{1-\mu}{\eta^2-\mu}}
\end{equation}
that separates stable from unstable flow \cite{Chandrasekhar1961} and implies a detachment of the unstable flow for any $\mu<0$. However, experiments and viscous calculations show that Taylor vortices extent beyond this neutral surface when counter-rotation sets in, see for example \cite{Taylor1923}. Esser and Grossmann \cite{Esser1996} deduced from their stability calculation that flow structures protrude the neutral surface by a factor $a(\eta)$, i.e. the effective extension of secondary flow is
\begin{equation}
 r_{EG}(\mu)=r_i+a(\eta) \, (r_n-r_i)
\end{equation}
with a factor
\begin{equation}
 a(\eta) = (1-\eta)\left[\sqrt{\frac{(1+\eta)^3}{2 (1+3\eta)}} -\eta\right]^{-1}
\end{equation}
that takes values between $1.4$ and $1.6$. The rotation ratio $\mu_{\mathrm{pred}}$ where the unstable flow detaches from the outer cylinder wall then follows from the condition $r_{EG}(\mu_{\mathrm{pred}})=r_o$ and reads
\begin{equation}
	\mu_{\mathrm{pred}}(\eta)=-\eta^2 \frac{(a^2-2a +1)\eta+a^2-1}{(2a-1)\eta+1} \, .
  \label{eq:mupred1}
\end{equation}
For $\mu<\mu_{\mathrm{pred}}$ the turbulence can no longer fill the whole cylinder gap, i.e. $r_{EG}<r_o$, and intermittency has to set in.
The extended range of the inner unstable region 
as captured by the factor $a(\eta)$ then mandates a minimal counter-rotation for the neutral surface to fall inside the cylinders.
The evaluation of \eqref{eq:mupred1} yields predictions \mbox{$\mu_{\mathrm{pred}}(0.5)=-0.191$} and \mbox{$\mu_{\mathrm{pred}}(0.71)=-0.344$} 
that compare well with the empirically found onsets of intermittency \eqref{eq:mucrit}, see also Tab. \ref{tab1}.

\begin{table}
 \begin{tabular}{cccccc}
	\toprule
	$\eta$ & $\mu_c$ & $\mu_{\mathrm{max}}$ & $\mu_\mathrm{pred}$ & $\mu_{bis}$ & $\mu_\mathrm{LSC}$\\
	\colrule
	$0.5$    & $-0.207$ & $-0.195$ & $-0.191$ & $-0.309$ & $-0.223$ \\
	$0.68$   &          & $-0.295$ & $-0.321$ & $-0.360$ & \\
	$0.71$   & $-0.351$ & $-0.357$ & $-0.344$ & $-0.367$ & $-0.357$ \\
	$0.716$  & $-0.368$ & $-0.33$  & $-0.349$ & $-0.368$ & \\
	$0.7245$ &          & $-0.333$ & $-0.356$ & $-0.370$ & \\
	\botrule
 \end{tabular}
 \caption{Rotation ratio of the onset of intermittency $\mu_c$ and of the torque maximum $\mu_{\mathrm{max}}$ together with  
	the new prediction $\mu_\mathrm{pred}$ \eqref{eq:mupred1} and the angle bisector $\mu_{bis}$ \eqref{eq:bisector} for various radius ratios. The rotation ratio $\mu_\mathrm{LSC}$ of the maximal mean-flow contribution to the torque is also given for the numerical simulations in the last column.}
 \label{tab1}
\end{table}

\section{Enhanced large scale circulation}
\label{sec:LSC}
We now turn to the question of why the torque maximum coincides with the onset of the radial intermittency. 
Van Gils et al. \cite{vanGils2012} argue that the torque decreases when turbulent bursting sets in because of the reduced radial transport. However, to obtain a torque maximum 
at the bursting onset $\mu_c$, additionally, the torque has to increase with $\mu$ decreasing from zero to $\mu_c<0$.
We argue that this increase is caused by a strengthening of the mean Taylor vortex flow. Such a large-scale circulation (LSC) is able to effectively transport momentum and, thus, to increase the torque in addition to turbulent fluctuations. The strengthening of the LSC is due to a change in the effective outer boundary condition: 
While for $\mu>\mu_{\mathrm{pred}}$ the LSC seeks to extend beyond the outer cylinder and is restricted by the rigid wall, i.e. $r_{EG}>r_o$, 
the rigid boundary conditions become replaced by a softer free-surface-like laminar outer boundary layer as $\mu$ approaches $\mu_{\mathrm{pred}}\sim\mu_c$. 
Less restricted, the large-scale vortices can become stronger before they will be destroyed by the bursting for $\mu<\mu_c$. 

\begin{figure} 
  \centering
  \includegraphics{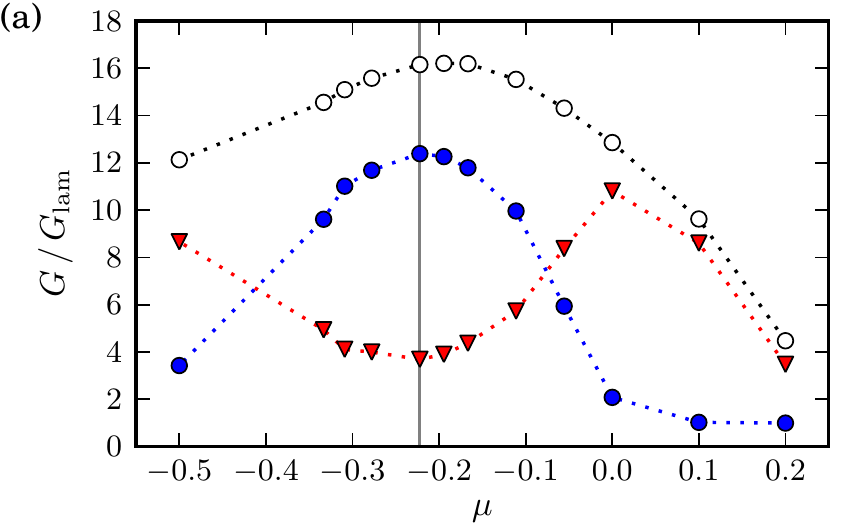}		
  \includegraphics{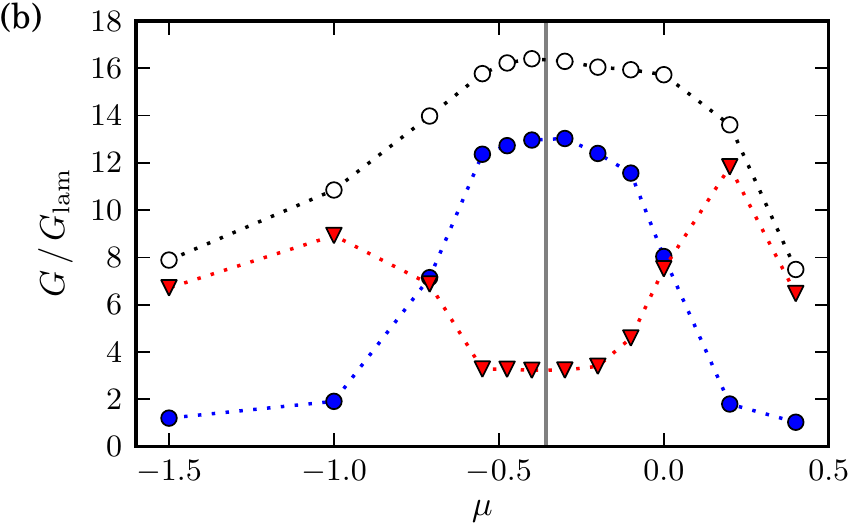}		
 \caption[]{Rotation dependence of the mean-flow (blue circles) and turbulent contribution (red triangles) to the torque (open circles).  
	The torque due to the mean flow exhibits a maximum marked by the vertical gray line. Dotted lines connecting the 
	data serve as guide to the eye. 
	(a) $\eta=0.5$ and $Re_S=20000$. (b) $\eta=0.71$ and $Re_S=19737$. 
  \label{fig:nupart}}
\end{figure}
To test this picture we decompose the flow into the LSC contribution $\bar{\bv{u}}=\left<\bv{u}\right>_{\varphi,t}$ 
that contains the mean variations in radial and axial direction and the turbulent fluctuations $\bv{u}'=\bv{u}-\bar{\bv{u}}$. 
With this decomposition one finds from Eq. \eqref{eq:Jw} a partitioning of the torque
\begin{equation}
 G = \bar{G} + G'
\end{equation}
with the mean-flow (LSC) and turbulent contribution
\begin{eqnarray}
 \bar{G} &=& \nu^{-2} \left< r^3\left(\left<\bar{u}_r\bar{\omega}\right>_{A(r),t}-\nu \partial_r\left<\bar{\omega}\right>_{A(r),t} \right)  \right>_r \, , \nonumber \\
 G' &=& \nu^{-2} \left< r^3\left<u_r'\omega'\right>_{A(r),t} \right>_r \, .
 \label{eq:G_contrib}
\end{eqnarray}
The mixed terms $\left<\bar{u}_r\omega'\right>$ and $\left<u_r'\bar{\omega}\right>$ in \eqref{eq:Jw} 
as well as $\left<\omega'\right>_{A(r),t}$ vanish due to the definition of $\bar{\bv{u}}$. 
While the complete torque $G$ is radially independent, similar expressions for 
mean-flow and turbulent contribution vary with the radius. Therefore, we introduced 
an additional radial average in \eqref{eq:G_contrib} to measure the mean weight of 
each contribution. Moreover to accurately capture the mean Taylor vortex motion, 
we axially shift the instantaneous flow fields during the temporal average of 
$\bar{\bv{u}}$ and in \eqref{eq:G_contrib} so that Taylor vortices always stay at a fixed height.

For $\mu\gtrsim0$, the torque is mainly caused by turbulent fluctuations and the mean-flow contribution 
nearly drops to the laminar level, as shown in Fig. \ref{fig:nupart}. Turbulent fluctuations also dominate 
the torque for strong counter-rotation, i.e. $\mu=0.5$ for $\eta=0.5$ and $\mu\lesssim0.71$ for $\eta=0.71$.
For intermediate rotation ratios, mean-flow vortices (LSC) contribute the major share to the torque. Note that
the onset of mean vortical flow for $\mu>-\eta$, with $\mu=-\eta$ corresponding to perfect counter-rotation 
$Re_o=-Re_i$, was previously observed by Ravelet et al. \cite{Ravelet2010}.
Additionally, the LSC contribution grows with $\mu$ decreasing from zero which is consistent with our 
picture of a change in the outer boundary condition from no-slip to a less restrictive free-surface condition.
The mean Taylor vortices are strongest (as measured by their contribution to the torque)
at the rotation ratio $\mu_{\mathrm{LSC}}(\eta)$ where $\bar{G}$ is maximized. 
Using a quadratic fit with
an uncertainty estimation analogue to \eqref{eq:uncert} we find
\begin{eqnarray}
 \mu_{\mathrm{LSC}}(0.5) &=& -0.223 \pm 0.018  \nonumber \\
 \mu_{\mathrm{LSC}}(0.71) &=& -0.357 \pm 0.075 \, .
\end{eqnarray}
Consequently, the rotation ratio of optimal momentum transport by the mean flow coincides with the 
empirically found onset of intermittency, Eq. \eqref{eq:mucrit}, 
within the given uncertainties. Furthermore, the mean-flow contribution is responsible for the maximum in 
the total torques, cf. Fig. \ref{fig:nupart} and Eq. \eqref{eq:mumax}, thereby establishing the 
connection between the onset of intermittency and the torque maximum within the framework depicted above.

\begin{figure} 
  \centering
  \includegraphics{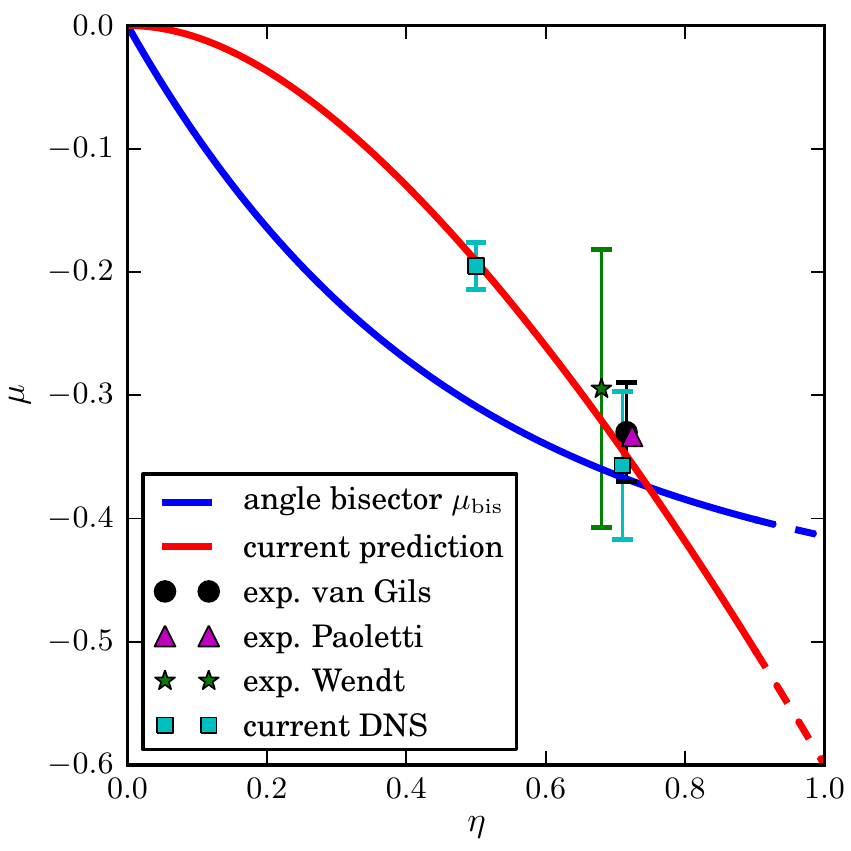}		
 \caption[]{Location of torque maxima for different radius ratios $\eta$. 
  Blue: Angle bisector \eqref{eq:bisector}.
	Red: boundary layer estimate \eqref{eq:mupred1}.
	The black circle, purple triangle and green star show torque maxima from the experiments \cite{vanGils2012},
	\cite{Paoletti2011} and \cite{Wendt1933}. The cyan squares mark the simulation result \eqref{eq:mumax}.
  \label{fig:prediction}}
\end{figure}
This connection implies that the prediction for the intermittency onset also acts as a prediction for torque maxima. 
We thus compare the predictions from the 
boundary layer argument $\mu_{\mathrm{pred}}(\eta)$ from Eq. (12) 
and the bisection argument $\mu_{\mathrm{bis}}(\eta)$ from Eq. (8)
\cite{vanGils2012} with experimental and numerical results for torque maxima in Fig. \ref{fig:prediction} (and Tab. \ref{tab1}). 
The rotation ratio $\mu=-2.797$ of the turbulent bursts found by Coughlin and Marcus \cite{Coughlin1996} lies 
below both predictions and, thus, clearly in the intermittent range.
Our simulation result for $\eta=0.71$ and Wendt's experimental result for $\eta=0.680$ are consistent with both the angle bisector and the current prediction within the error bars. 
Moreover, we note that the torque maximum \mbox{$\mu_{\mathrm{max}}(0.7245)=-0.333$} measured by Paoletti and Lathrop 
\cite{Paoletti2011} as well as \mbox{$\mu_{\mathrm{max}}(0.716)=-0.33\pm0.04$} measured by van Gils et al. \cite{vanGils2012} 
tend towards our prediction, Eq. \eqref{eq:mupred1}. However considering the usual error bars, these values are also consistent with the angle bisector line.
The simulation results for $\eta=0.5$ provide a better test of both 
predictions, since the values obtained from (8) and (12) differ. The numerical data
are in better agreement with the boundary layer estimate (12).

\section{Final remarks}
\label{sec:remarks}
The analysis presented here support the idea that the torque increases with increasing counter-rotation because the vortices gain in strength until they can no longer sustain turbulence all across the gap. The torque drops for stronger counter-rotation, when the 
detachment of mean vortices from the outer layer leads to radial intermittency.

The predictions from the boundary layer argument presented here and the angle bisection
proposal of van Gils et al. \cite{vanGils2012} give indistinguishable predictions for a radius ratio of $\eta\approx0.75$, but the shape of the $\eta$-dependencies 
is sufficiently different that data in particular for smaller $\eta$ should allow to 
distinguish between the two. The available data for $\eta=0.5$ are in good
agreement with the present argument. 
Clearly, results for more $\eta$ are required and work along those lines is in progress.

For larger $\eta$ in the limit $\eta\rightarrow 1$ 
the current theory predicts a maximum for a rotation ratio of $-0.6$ ,
whereas the angle bisection gives a value close to $-0.4$. However, this limit is 
delicate because the linear instability disappears and both theories will most 
likely have to be refined or replaced. Evidence for this is 
provided for instance by the measurements by Ravelet et al. 
\cite{Ravelet2010} for $\eta=0.917$, which do not show a torque maximum for counter-rotation. Figure \ref{fig:prediction} reflects this uncertainty in the prediction by the 
change from a continuous to a dashed line for $\eta>0.9$.

Furthermore, numerical simulations revealed that at lower $Re_S\leq4\cdot10^3$ the torque is maximized at $\mu\approx0$ 
for $\eta=0.71$ \cite{Brauckmann}. This larger rotation ratio is not covered with the current theory, so that further refinements are needed for lower Reynolds numbers and
mildly turbulent flows.

\begin{acknowledgements}
We are grateful to Marc Avila for developing and providing the code used for our
simulations. This work was supported in part by Deutsche Forschungsgemeinschaft. 
Most computations were done at the LOEWE-CSC in Frankfurt.
\end{acknowledgements}

\appendix
\section{Re-analysis of Wendts data}
\label{sec:wendt}
\begin{figure} 
  \centering
  \includegraphics{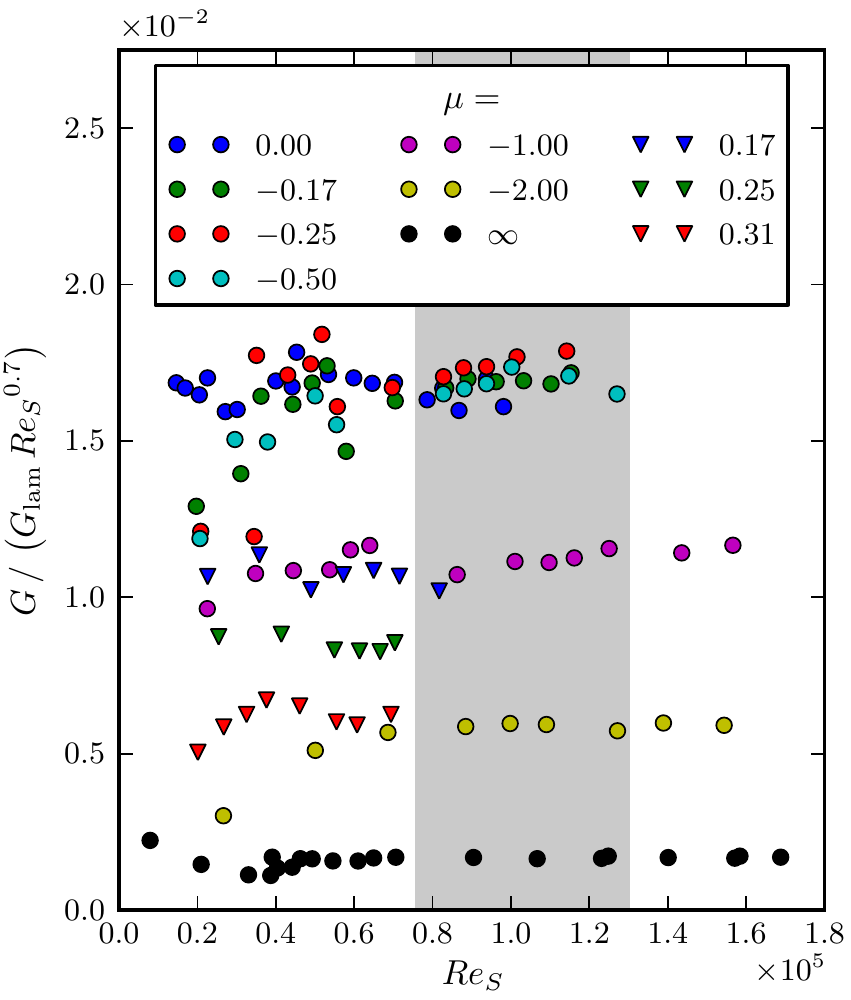}		
 \caption[]{Torques measured by Wendt \cite{Wendt1933} for $\eta=0.680$ and various rotation ratios. The values are compensated with the laminar $G_{\mathrm{lam}}$ and with the effective scaling $G/G_{\mathrm{lam}}\sim {Re_S}^{0.7}$ reported by Wendt. The torques for the range shaded in gray are further analyzed in Fig. \ref{fig:max-wendt} . 
  \label{fig:nu-wendt}}
\end{figure}
Recent experimental studies analyze the dependence of torque on the shear rate and on the mean system rotation independently. This decomposition is advantageous since torques can be compensated either by dividing by the effective scaling with the shear \cite{vanGils2011} or by talking the ratio to $G(\mu=0)$ \cite{Paoletti2011} to study the rotation dependence. The resulting torque amplitudes are based on numerous measurements for each rotation ratio which improves statistical significance. In contrast, Wendt presented the torque-dependence on the rotation for some selected shear Reynolds numbers in figure 10 of Ref. \cite{Wendt1933}. Since this evaluation is based on single measurements, uncertainties may play a major role.

\begin{figure} 
  \centering
  \includegraphics{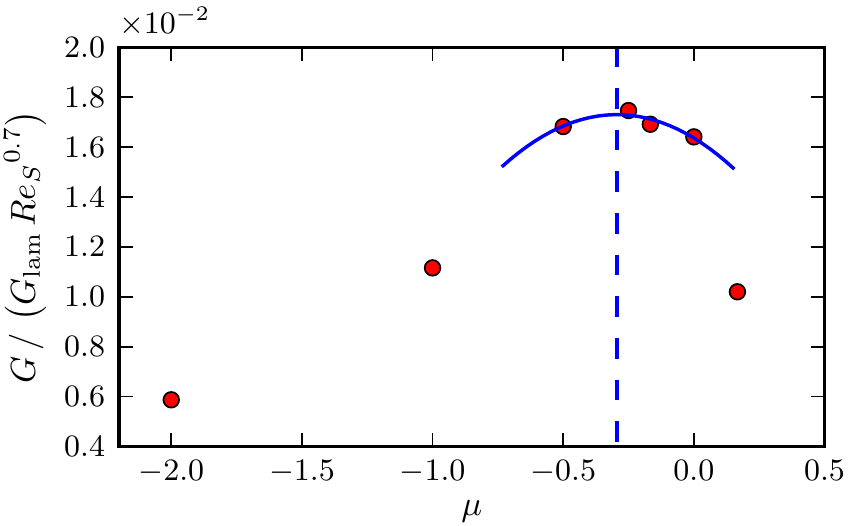}		
 \caption[]{Compensated torques by Wendt for $\eta=0.680$ averaged in the range $7.8\e{4}<Re_S<1.3\e{5}$ for each rotation ratio independently. The blue line indicates a quadratic least-square fit to the four largest values. Its maximum $\mu_{\mathrm{max}}=-0.295$ is marked by the dashed line.
  \label{fig:max-wendt}}
\end{figure}
Therefore, we here apply the current analysis method to Wendt's torque measurements for $\eta=0.680$ digitized from figure 9 in \cite{Wendt1933}. Figure \ref{fig:nu-wendt} shows the torques for various rotation ratios compensated with ${Re_S}^{0.7}$ which Wendt found as effective scaling for $10^4\lesssim Re_s\lesssim 10^5$. One easily sees that the torque depends on the mean rotation with the largest values for high $Re_S$ at $\mu=-0.25$. 
We closely follow the analysis in \cite{Paoletti2011,vanGils2011,vanGils2012} and average the compensated torques in the range $7.8\e{4}<Re_S<1.3\e{5}$ to find amplitudes depending on the rotation only, see Fig. \ref{fig:max-wendt}. 
We chose a Reynolds number range that starts after the shift of the torque maximum \cite{Brauckmann} and includes the highest data points for $-0.50\leq\mu\leq-0.17$ (cf. Fig. \ref{fig:nu-wendt}). 
One observes a maximum in the statistical more significant mean amplitudes for moderate counter-rotation which was also found in recent studies \cite{Paoletti2011,vanGils2011,vanGils2012} and in current simulations. 
Based on a quadratic fit to the largest amplitudes we find
\begin{equation}
 \mu_{\mathrm{max}}(0.680)= -0.295 \pm 0.113 \, ,
\end{equation}
with the uncertainty calculated in analogy to Eq. \eqref{eq:uncert}. Its relative high level is due to the broad maximum in Fig. \ref{fig:max-wendt} and due to the few rotation ratios investigated by Wendt. In spite of the high uncertainty, the torque maximization for counter-rotation, i.e. $\mu_{\mathrm{max}}<0$, is clear without ambiguity. Moreover, the new maximum, $\mu_{\mathrm{max}}(0.680)=-0.295$, lies consistently between the maxima identified here, cf. Eq. \eqref{eq:mumax}.


\end{document}